\documentclass[aps,fleqn,twocolumn]{revtex4}
 
\usepackage{graphicx}
\usepackage{amsmath} 
\usepackage[charter]{mathdesign}

\def\d{{\rm d}}
\def\e#1{{\rm e}^{#1}}

\def\x{r}
\def\xq{\alpha}
\def\pe{p\hspace{-0.08em}_{E}}

\def\sd{s}
\def\taust{\tau_{\!{\rm s}}}
\def\tauze{\tau_{\!{\rm z}}}

\def\del#1{\delta\hspace{-0.08em}#1}
\def\chis{\chi_{\rm S}}
\def\chic{\chi_{\rm C}}

\begin{document}
\title{Proper time delays measured by optical streaking}
\author{Ulf Saalmann}
\author{Jan M. Rost}
\affiliation{Max-Planck-Institut f{\"u}r Physik komplexer Systeme,
 N{\"o}thnitzer Str.\ 38, 01187 Dresden, Germany}
\date{\today}

\begin{abstract}\noindent
The generation of a streaking spectrogram is based on energy absorption from the streaking laser.
Investigating this absorption we show rigorously under which condition the measured time shift is independent of properties of the streaking light. 
In this case it provides the Wigner-Smith time delay.
The latter is infinite for systems with a long-range potential tail, such as Coulomb systems. 
Here, we suggest to determine the time delay relative to pure hydrogen for meaningful results. 
Finite delays obtained so far for Coulombic systems without the hydrogen reference are the consequence of a finite streaking frequency and depend on its value as well as on the electron's excess energy.
Our analysis also suggests a time-delay measurement technique that avoids the record of a complete streaking scan.
\end{abstract}

\maketitle
\noindent
Attosecond laser pulses can in principle determine ultrashort time spans to uncover microscopic details of dynamical processes.  
To date, however, it is technically challenging to produce two strong enough attosecond pulses for a pump-probe experiment, i.\,e., to start and stop the clock. 
One way to bypass this obstacle is the so-called streaking method \cite{cota+97}: An attosecond laser pulse starts the clock by emitting an electron. A second, weak near-infrared pulse, phase-locked with a tunable delay $s$ to the attosecond pulse, ``streaks'' this photo-electron, i.\,e. influences its momentum in the continuum while it is leaving its (binding) potential.
From the streaking spectrogram, the energy (or momentum) distribution as a function of $s$, one can extract so-called streaking delays $\taust$.
These delays differ for photo-electrons coming from different orbitals, as revealed in seminal experiments \cite{camu+07,scfi+10}.
In the case of atoms \cite{scfi+10}, these delays have been linked to the Wigner-Smith time delay $\del{t}$ from scattering theory \cite{wi55,sm60}, revealing how much time a particle spends traversing a potential in comparison with a free particle with the same energy $E$.
Comparison of streaking delays with those delays has provided theoretical evidence that both agree for short-range potentials and typical laser parameters used \cite{pana+15}. 
Nevertheless, a direct extraction of the Wigner-Smith time delay from the energy, absorbed by the photo-electron from the streaking laser, is lacking. 

Here we provide an analytical derivation of this connection. 
It shows that a meaningful extraction is possible if the streaking laser fulfills specific conditions and it also explains the confusing fact that experiments measure finite streaking delays for photo-electrons, i.\,e., Coulombic systems with long-range potentials, although for such potentials the Wigner-Smith time delay is infinite.

Basically all theoretical accounts of the streaking technique (see various reviews \cite{dalh+12,maca+14,pana+15}) rely on the strong-field approximation (SFA),
which neglects the potential for the continuum electron although the field-potential interaction is the source of energy absorption from the streaking laser. 
Moreover, streaking measurements are ``classical clocks'' \cite{pana+15}. Other than ``quantum clocks'' measuring time delays \cite{klda+11,dagu+12,issq+17}, they
 do not rely on quantum interferences. 
Classical streaking calculations have been shown to coincide with quantum calculations \cite{pana+13,pana+15}, as long as there are no resonances in the continuum \cite{issq+17}. Hence, we use classical dynamics in the following, which naturally allow us to include the potential and to work out the corner stones of streaking time delays from energy absorption clearly.
%%%%%%%%%%%%%%%%%
\begin{figure}[b]\centering
\includegraphics[width=0.7\columnwidth]{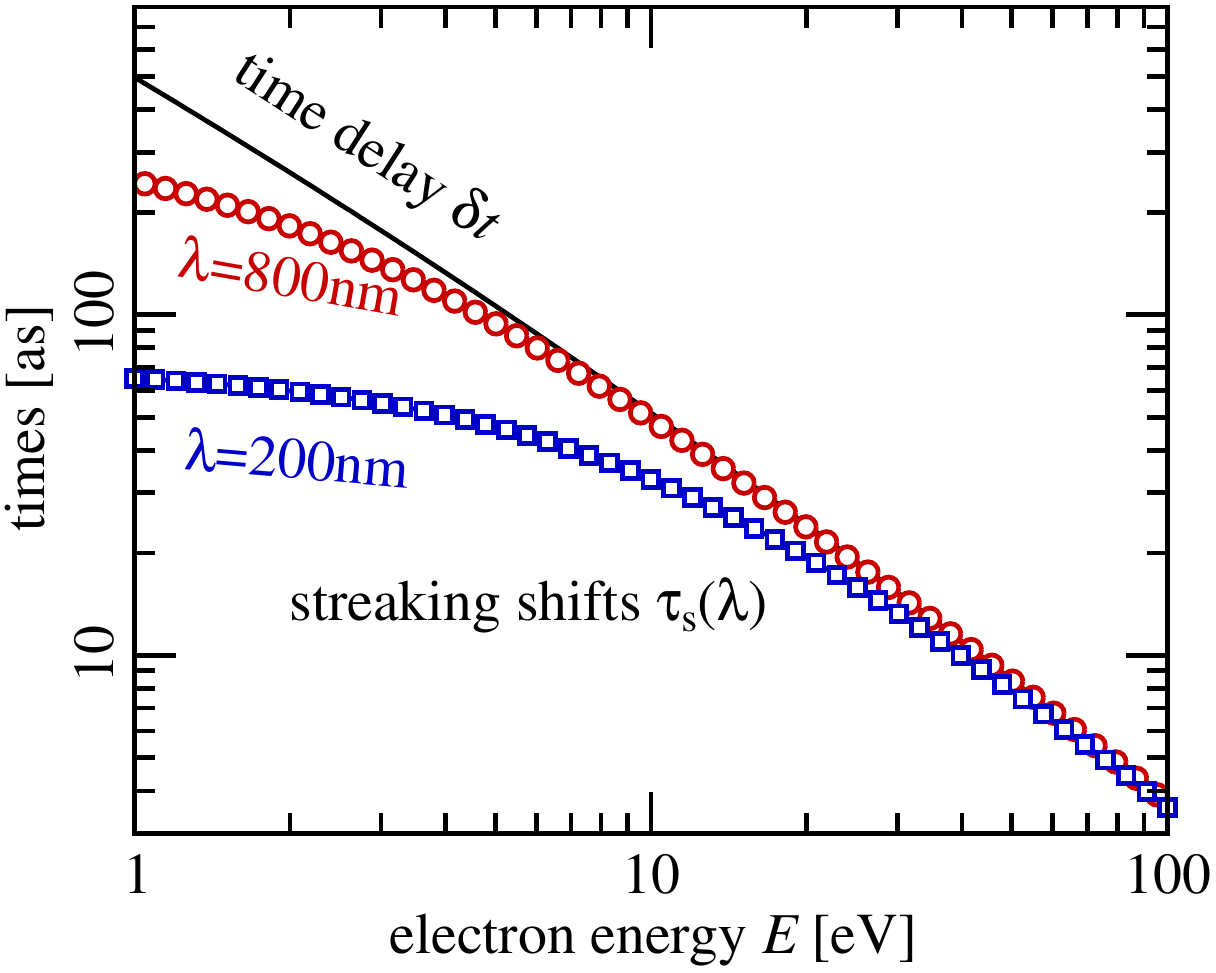}
\caption{Wigner-Smith time delay $\del{t}$ (black solid line) and streaking delays $\taust$ for two wavelengths, $\lambda\,{=}\,200$\,nm (blue squares) and $\lambda\,{=}\,800$\,nm (red circles) as function of the electron energy $E$ for a Yukawa potential $V(r)\,{=}\,\e{-r/r_{\rm yuk}}/r$ with $r_{\rm yuk}\,{=}\,3$\,\AA.}
\label{fig:yuk}
\end{figure}%
%%%%%%%%%%%%%

The time delay of a ``half collision'' in the spherical potential $V(\x)$ initiated at time $t\,{=}\,0$ with energy $E$ follows from the action difference \cite{ro98}
\begin{equation}
\label{eq:action}
S(E) = \int\limits_{0}^{\infty}\!\!\d r\,\big[p(r)-\pe\big]
\end{equation}
through the energy derivative $\del{t} ={\d}S(E)/{\d E}$ which yields
\begin{equation}
\label{eq:dt1}
\del{t} 
= \int\limits_{0}^{\infty}\!\!\d\x\,\,\bigg[\frac{1}{p(\x)}-\frac{1}{\pe}\bigg]
= \int\limits_{0}^{\infty}\!\!\d t\,\,\bigg[1-\frac{p(t)}{\pe}\bigg]\,,
\end{equation}
where in the last expression $\d\x\,{=}\,p\,\d t$ has been used.
The time delay $\del{t}$ measures how much more or less time it takes a particle to move in a potential $V$ with momentum $p(\x)\,{=}\,\sqrt{2[E{-}V(\x)]}$ compared to a free particle with momentum $\pe\,{=}\,\sqrt{2E}$. 
This is particularly obvious from the second equation in \eqref{eq:dt1}.
As noted by Smith \cite{sm60}, for $\del{t}$ to be well-defined, the particle must be asymptotically free. 

Figure~\ref{fig:yuk} shows $\del{t}$ for a short-range Yukawa potential (black line).
The time delay is compared to streaking shifts $\taust$, with a laser field of frequency $\omega$ and peak vector potential $A = F/\omega$, where the field strength $F$ is so small such that $A\ll \pe$.
The streaking field
\begin{equation}\label{eq:asin}
A_{\sd}(t)=-A\,\sin\big(\omega[t{+}\sd]\big),
\end{equation} 
can be shifted by a delay $\sd$ with respect to the particle's release at $t\,{=}\,0$,
often achieved by attosecond photo-ionization \cite{scfi+10,kigo+04,itqu+02,yaba+05,frwi+09,khiv10,zhth11,ivsm11,pana+13}. 
In a full streaking scan, $s$ is varied from ${-}T/2$ to ${+}T/2$, i.\,e., over the period $T\,{=}\,2\pi/\omega$ of the streaking laser.
We show in Fig.\,\ref{fig:yuk} streaking shifts $\taust$ for two different laser wavelengths $\lambda$. 
Obviously, the time delay $\del{t}$ agrees with those streaking shifts $\taust$ only for sufficiently large energies $E$.

In order to determine the general conditions under which streaking delays reveal the Wigner-Smith time delay, we investigate how the streaking field $A_{\sd}(t)$ changes the momentum of the electron while leaving the potential region. 
To 1st order in the vector potential $A$, the momentum change is given by
\begin{equation}\label{eq:deltap}
\del{p}(s) = A_{\rm s}(0) + \frac{1}{\pe}\int\limits_{0}^{\infty}\!\!\d t\,\big[A_{\rm s}(t){-}A_{\rm s}(0)\big]\,V'\big(r(t)\big)
\end{equation}
with the vector potential $A_{\rm s}(t)$ and all other quantities from the field-free ($A\,{=}\,0$) dynamics. 
The derivation of Eq.\,\eqref{eq:deltap} is straight-forward in the Kramers-Henneberger frame and requires only the perturbative character of the streaking field $A\,{\ll}\,\pe$.
Details can be found in the supplemental material \cite{suppl}. 
%%%%%%%%%%%%%%%%%
\begin{figure}[t]\centering
\includegraphics[width=\columnwidth]{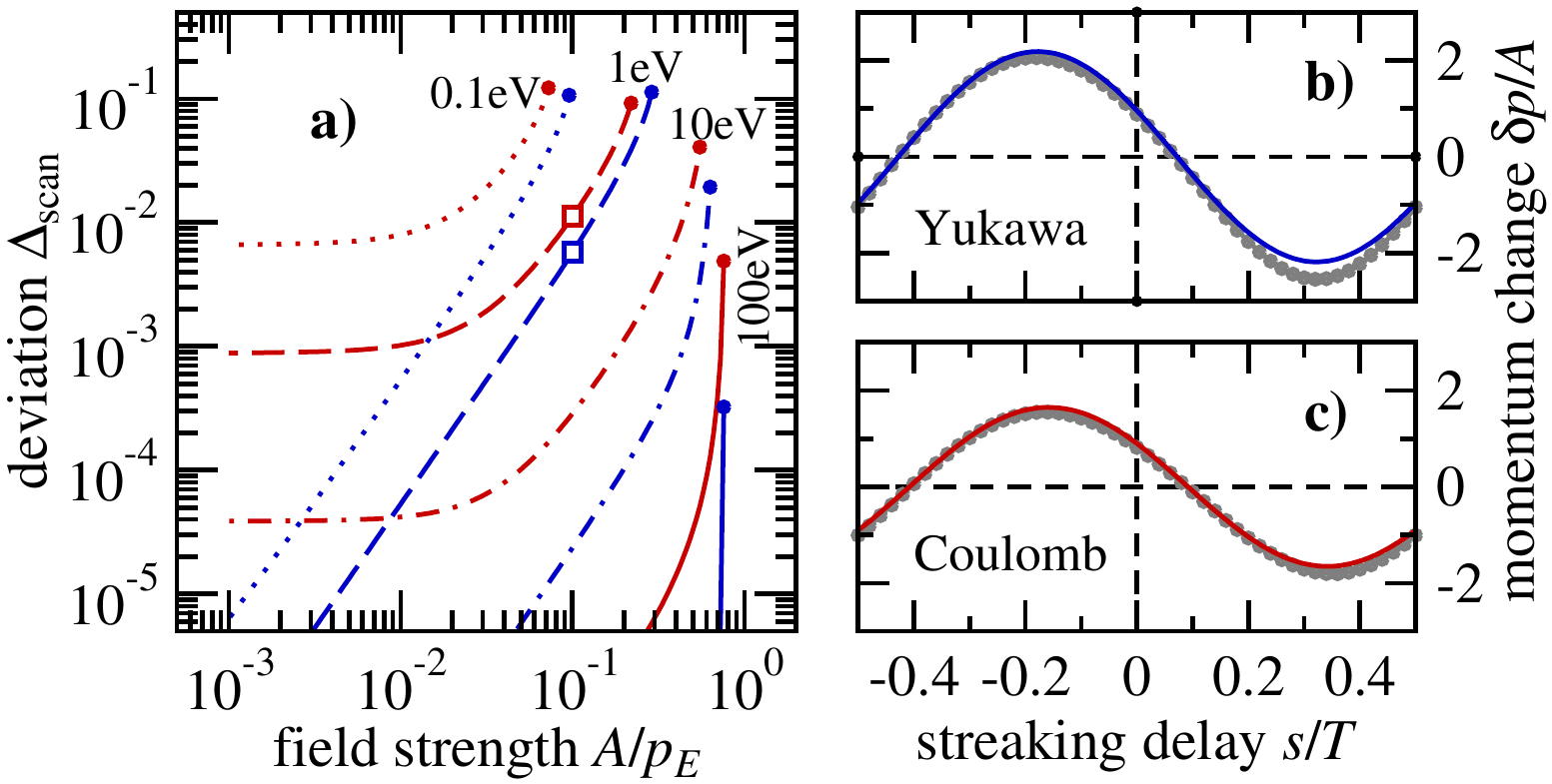}
\caption{Comparison of the harmonic streaking scans $\del{p}(\sd)$ from Eq.\,\eqref{eq:dpharm} with respect to a numerical solution. The relative deviation is defined as $\Delta_{\rm scan}\,{\equiv}{\int}\!\d\sd\,[\del{p}(\sd)\,{-}\,\del{p}_{\rm num}(\sd)]^{2}\big/{\int}\!\d\sd\,[\del{p}_{\rm num}(\sd)]^{2}$ and shown for four energies (\textbf{a}) as a function of the laser field strength $A$, with $A$ given in units of $\pe$. 
Result are for the Yukawa potential, as in Fig.\,\ref{fig:yuk}, (blue lines) and the Coulomb potential (red).
The circles mark the largest $F$ for which a full streaking scan can be made.
The squares mark the two examples, having relative large deviations $\Delta_{\rm scan}$, shown to the right (\textbf{b},\,\textbf{c}), where $\del{p}_{\rm num}$ (gray circles) is compared to the harmonic expression $\del{p}$ (blue/red line) from Eq.\,\eqref{eq:dpharm}.}
\label{fig:trace}
\end{figure}%
%%%%%%%%%%%%%

We have written \eqref{eq:deltap} in a form which makes the relation to the often-used SFA apparent. 
The latter neglects the potential $V$ leaving the well-known result that $\del{p}(s) = A_{s}(0)$, the vector potential at release time $t{=}0$.
We would like to stress at this point that all results below are based on exact numerical calculations, we use Eq.\,\eqref{eq:deltap} for interpretation only. 
Figure~\ref{fig:trace} shows, however, that it gives even quantitatively correct results for sufficiently small streaking fields $F$ or $A$, respectively.

Note that Eq.\,\eqref{eq:deltap} is valid for arbitrary forms $A_{\rm s}(t)$. 
For the harmonic streaking field \eqref{eq:asin} it has the particularly compact form
\begin{subequations}\label{eq:dpharm}\begin{align}
\label{eq:dpharm1}
\del{p}(\sd) & = -\frac{F}{\omega}\big[\chis\sin(\omega\sd)+\chic\cos(\omega\sd)\big]
\\
\label{eq:dpharm2}
\chis & \equiv 1-\frac{1}{\pe}\int\limits_{0}^{\infty}\!\!\d t\,[1{-}\cos(\omega t)]\,V'\big(r(t)\big)
\\
\label{eq:dpharm3}
\chic & \equiv \frac{1}{\pe}\int\limits_{0}^{\infty}\!\!\d t\,\sin(\omega t)\,V'\big(r(t)\big),
\end{align}\end{subequations}
which reveals that $\del{p}$ is harmonic as well, with amplitude (in units of the quiver momentum $F/\omega$) and phase defined by the newly introduced quantities $\chis$ and $\chic$. 

With the momentum change \eqref{eq:dpharm} we can elucidate the relation to the Wigner-Smith time delay
\eqref{eq:dt1}, which can be written (using integration by parts) as
\begin{equation}
\label{eq:dt2}
\del{t} = -\frac{1}{\pe}\int\limits_{0}^{\infty}\!\!\d t\,t\,V'\big(\x(t)\big).
\end{equation}
Without loss of generality we may compare \eqref{eq:dt2} to the streaking-induced momentum change at $s\,{=}\,0$ from \eqref{eq:dpharm}
\begin{equation}
\del{p}(0)=-\frac{F}{\omega}\chic
=-F\int\limits_{0}^{\infty}\!\!\d t\,\frac{\sin(\omega t)}{\omega}\,V'\big(r(t)\big)\,.
\end{equation}
It is obvious that for small streaking frequencies $\omega$,
$\del{p}(0)$ directly gives the Wigner-Smith time delay
\begin{equation}\label{eq:dtdp}
\tauze\equiv\frac{\del{p}(0)}{F}=\frac{\chic}{\omega}
\quad\mbox{with}\quad
\del{t}=\lim_{\omega\to0}\tauze.
\end{equation}
Since $\del{p}$ is measurable, Eq.\,\eqref{eq:dtdp} provides a direct way to determine $\del{t}$ experimentally, provided that the streaking frequency $\omega$ is sufficiently small and that the streaking vector potential $A$ is sufficiently weak to allow for a 1st-order description of the momentum change or energy absorption, cf.~Fig.\,\ref{fig:trace}.
Note, that there is no need for a full streaking scan of many delays $\sd$, a single measurement with $\sd\,{=}\,0$ contains the necessary information. 

Turning to traditional streaking measurements, the momentum change \eqref{eq:dpharm} may be written as
\begin{subequations}\label{eq:dpharmb}\begin{align}
\del{p}(\sd) & = -A \,\sqrt{\chis{\!}^{2}{+}\chic{\!}^{2}}\,\sin\big(\omega[\sd\,{+}\,\taust]\big)
\\
\label{eq:dpharmb2}
\taust & \equiv \frac{1}{\omega}\arctan(\chic,\chis),
\end{align}\end{subequations}
which is, as Eq.\,\eqref{eq:dpharm1}, a harmonic function of $\sd$ with a displacement in the phase given by the streaking shift $\taust$.
Both times $\tauze$ and $\taust$ are, for sufficiently weak lasers, independent of the field strength $F$, as observed experimentally \cite{scfi+10}.
From $\lim_{\omega{\to}0}\chis=1$ and $\lim_{\omega{\to}0}\chic=\omega\,\del{t}$, cf.\ Eq.\,\eqref{eq:dtdp}, follows immediately
\begin{equation}\label{eq:taustlim}
\lim_{\omega{\to}0}\taust=\lim_{\omega{\to}0}\frac{\arctan(\omega\,\del{t},1)}{\omega}=\del{t}.
\end{equation}
Equations~\eqref{eq:dtdp} and \eqref{eq:taustlim} are the main result of this work, since they show that both single and full-scan streaking measurements may indeed provide the Wigner-Smith time delay.
%%%%%%%%%%%%%%%%%
\begin{figure}[b]\centering
\includegraphics[width=\columnwidth]{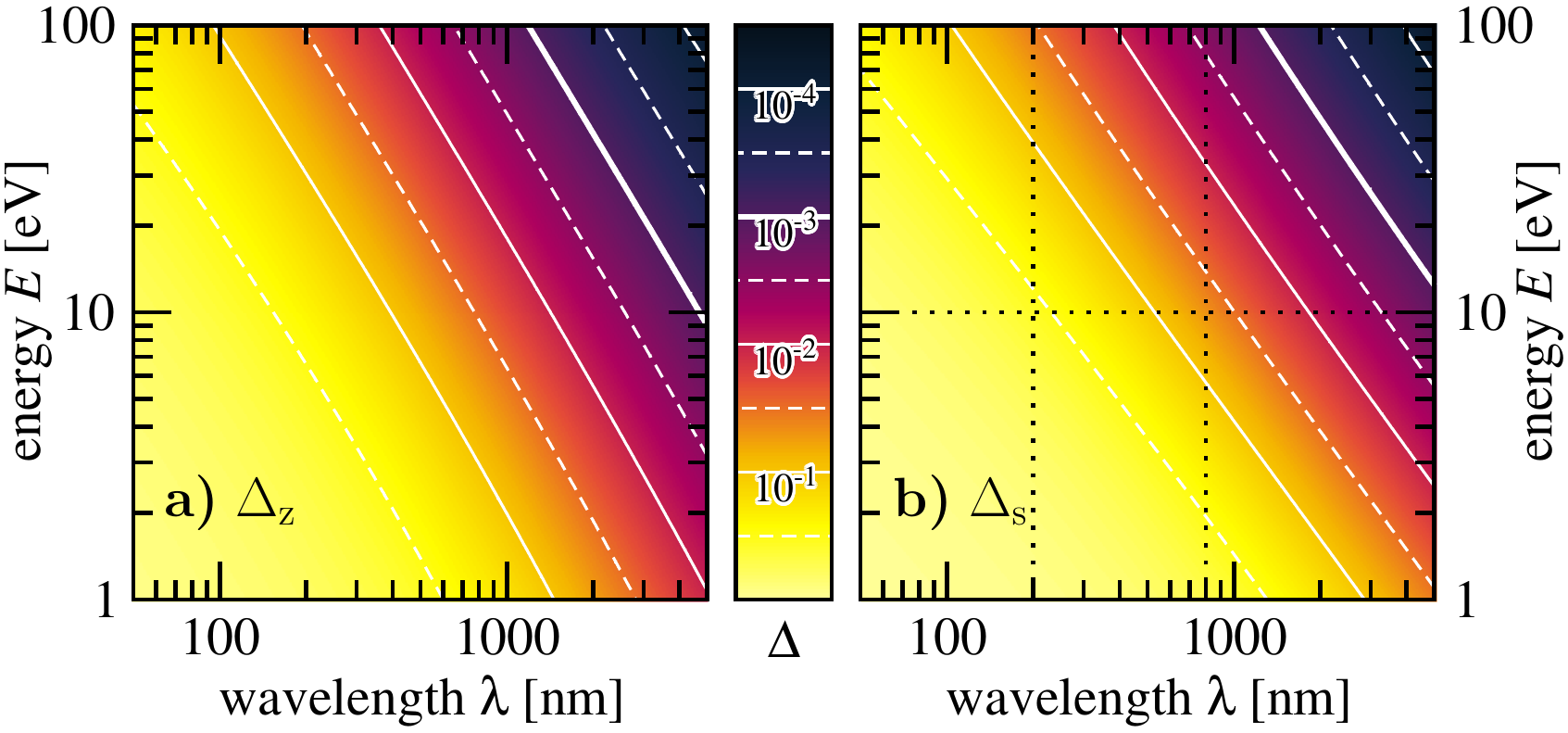}
\caption{Deviation of the zero-streaking time $\tauze= \del{p}(0)/F$ according to \eqref{eq:dtdp} and the full-streaking time $\taust$ according to \eqref{eq:dpharmb2} from the Wigner-Smith delay $\del{t}$ for the Yukawa potential used before in Fig.\,\ref{fig:yuk}.
Shown are the relative errors $\Delta_{\rm z}\,{\equiv}\,|\tauze{-}\del{t}|/\del{t}$ 
and $\Delta_{\rm s}\,{\equiv}\,|\taust{-}\del{t}|/\del{t}$ as a function laser wavelength $\lambda$ and electron energy $E$.
The vertical dotted lines mark the two wavelengths already shown in Fig.\,\ref{fig:yuk}, the horizontal one the energy shown in Fig.\,\ref{fig:coul}a.
The thicker white lines are the lower bounds for 1\textperthousand\ accuracy and can be approximately described by
$E\,{>}\,135\mbox{eV}\,[\lambda/1\mu\mbox{m}]^{-5/3}$
and
$E\,{>}\,126\mbox{eV}\,[\lambda/1\mu\mbox{m}]^{-3/2}$,
respectively.}
\label{fig:freq}
\end{figure}%
%%%%%%%%%%%%%

To be quantitative, we investigate next, under which conditions
the streaking delays $\taust$ and $\tauze$ really give the Wigner-Smith time delays $\del{t}$ for a Yukawa potential, which has been used as prototypical short-range potential in this context \cite{pana+13}. 
Figure \ref{fig:freq} provides the evolution of the relative differences
 $\Delta_{\rm z}\,{\equiv}\,|\tauze\,{-}\,\del{t}|/\del{t}$ 
and $\Delta_{\rm s}\,{\equiv}\,|\taust\,{-}\,\del{t}|/\del{t}$ as a function of electron energy $E$ and streaking laser wavelength $\lambda$. Indeed, for ``typical'' energies $E\,{\gg}\,10$\,eV and Ti:Sapphire laser pulses ($\lambda\,{=}\,800$\,nm) there is reasonable agreement between streaking shifts $\taust$, $\tauze$ and time delays $\del{t}$, as reported before \cite{pana+13,pana+15}. However, this agreement is by no means generally guaranteed\,---\,it requires, depending on the energy $E$, sufficiently large wavelengths $\lambda$ in agreement with Eqs.~(\ref{eq:dtdp}) and \eqref{eq:taustlim}.
While the overall behavior of both streaking delays is similar, one can see that $\Delta_{\rm s}$ grows faster than $\Delta_{\rm z}$.
Details regarding location and shape of the contour lines depend, of course, on the underlying potential. However, larger energies and longer wavelengths are always an option for short-range potentials to reach the parameter regime where the streaking time delays approach  $\del{t}$. 

\begin{figure}[t]\centering
\includegraphics[width=0.9\columnwidth]{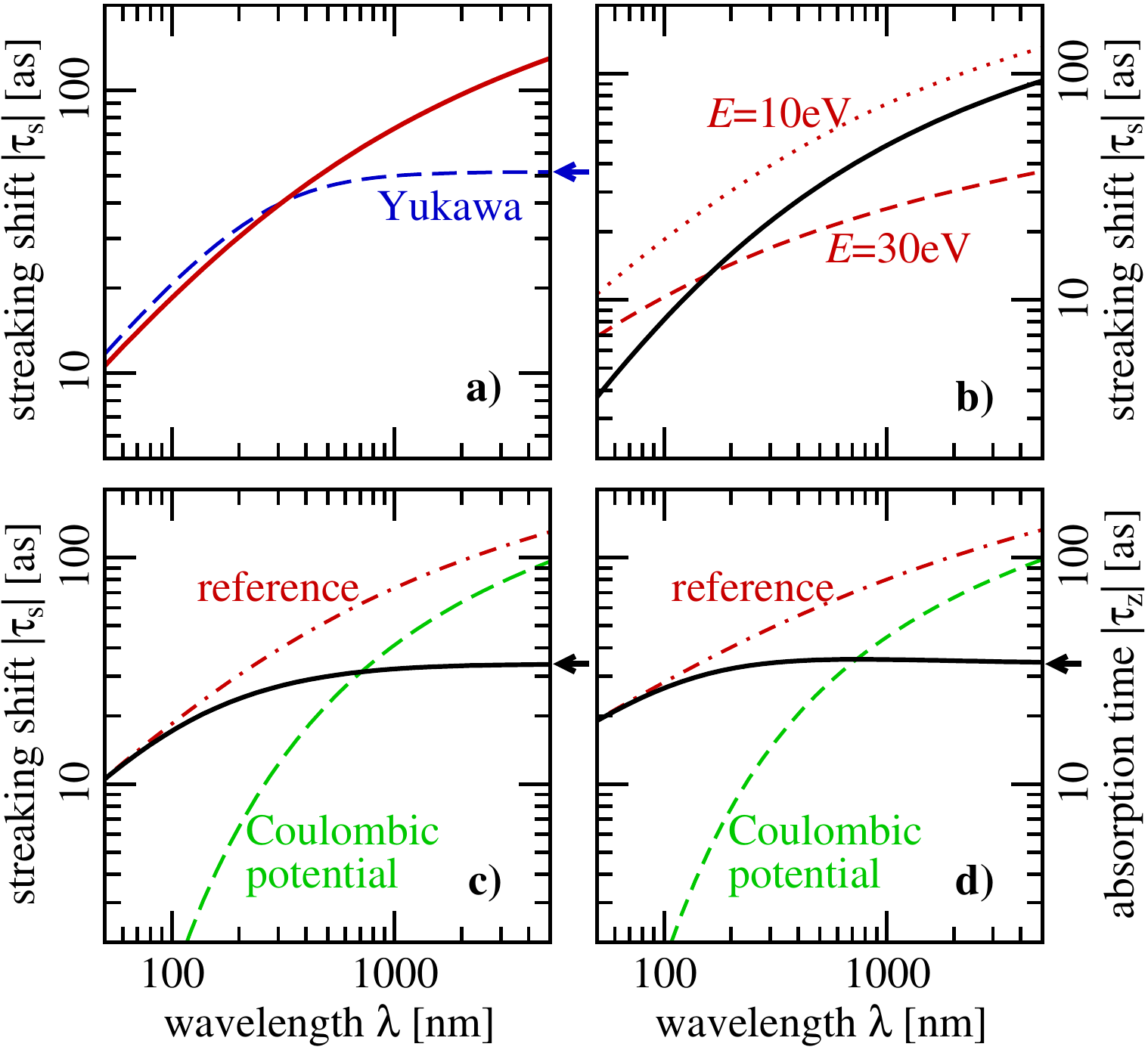}
\caption{Streaking shifts $\taust$ (and $\tauze$) for Coulomb potentials as a function of the streaking laser wavelength $\lambda$. In order to allow for a logarithmic scale we show absolute values for all times.
\textbf{a)} Comparison of Coulomb (red solid lines) and Yukawa, as in Fig.\,\ref{fig:yuk}, (blue dashed) potential for $E\,{=}\,10$\,eV. 
\textbf{b)} Shifts for two different energies (red dotted/dashed) and their respective difference (black solid). 
\textbf{c,\,d)} $\taust$ and $\tauze$ for the smoothed Coulomb potential $V(r)\,{=}\,{-}1/\sqrt{r^{2}{+}r_{\rm sm}{\!}^{2}}$ with $r_{\rm sm}\,{=}\,3$\,\AA\ (green dashed) and the newly defined $\tau_{\rm c}$ (black solid) using the Coulomb potential (red dot-dashed) as reference, both obtained for the same energy $E\,{=}\,10$\,eV.
The arrows at the right give values for $\lambda\,{=}\,100\,\mu$m, indicating converged values.}
\label{fig:coul}
\end{figure}

This, however, does not hold any longer for long-range potentials, most importantly, for the Coulomb potential. 
Since $\del{t}$ diverges and the streaking delays $\tauze$ and $\taust$ agree with $\del{t}$ for $\omega\,{\to}\,0$ or $\lambda\,{\to}\,\infty$, they should also diverge in this limit and in fact they do. 
Yet, the situation is more intricate as streaking for larger $\omega$ is only sensitive to a certain part of the potential (notably the inner part as shown below) and therefore leads to a \emph{finite\/} momentum change.
This is, however, not only a property of the Coulomb potential itself but depends sensitively on the parameters used, namely electron energy and streaking frequency, as illustrated in Fig.~\ref{fig:coul}a. In other words, these streaking time delays, although finite, are not suitable to characterize a Coulombic system.

%%%%%%%%%%%%%%%%%
%%%%%%%%%%%%%%
 
Rather, one should take the pure (hydrogen's) Coulomb potential $V_{\rm ref}(r)\,{=}\,{-}1/r$ as a reference (instead of the free particle) to determine time delays of systems with a long-range Coulomb tail,
an idea that has been put forward in the context of RABBIT \cite{dalh+12}. 
That means we define a Coulomb action $S^{\rm c}(E)$ by replacing in Eq.\,\eqref{eq:action} the free-particle momentum $\pe$ by $p_{\rm ref}(r)\,{=}\,\sqrt{2[E{-}V_{\rm ref}(\x)]}$ or, equivalently, subtract the action of the hydrogen reference from the actual Coulomb-tailed system. 
We obtain a finite time delay 
 \begin{equation}
\label{eq:dtc}
\del{t}^{\rm c} \equiv \frac{\d S^{\rm c}(E)}{\d E} = \frac{\d S(E)}{\d E}-
\frac{\d S_{\rm ref}(E)}{\d E}\,.
\end{equation}
 In analogy to \eqref{eq:dtc} we define
\begin{equation}\label{eq:taucoul}
\tau_{\!\sigma}{\!}^{\rm c} \equiv \tau_{\!\sigma}-\tau_{\!\sigma}{\!}^{\rm ref}
\quad\mbox{with}\quad \sigma={\rm s, z}
\end{equation}
whereby the reference times obtained for $V_{\rm ref}$ introduced in \eqref{eq:dtc} above.
Note that in Eq.\,\eqref{eq:taucoul} the individual terms diverge for $\omega\,{\to}\,0$, the difference, however, has a finite limit, independent of the properties of the streaking field if it is sufficiently weak, as illustrated with 
Figs.~\ref{fig:coul}c and \ref{fig:coul}d. This limit establishes the proper streaking time delays for potentials with a Coulomb tail.
%%%%%%%%%%%%%%%%%
\begin{figure}[t]\centering
\includegraphics[width=\columnwidth]{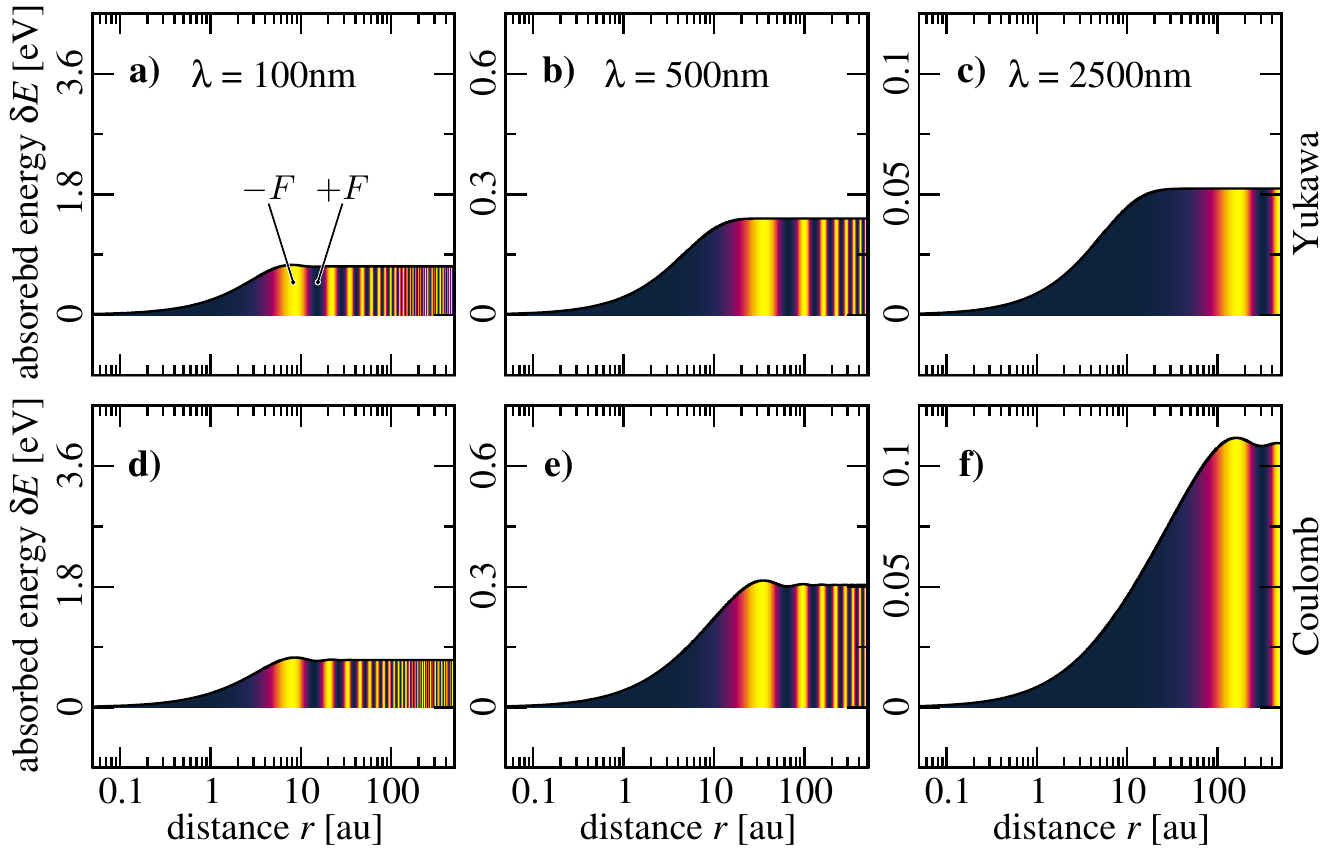}
\caption{Absorbed energy $\del{E}(r)\,{=}\,E(r)\,{-}\,E$ as a function of distance $r$ for the Yukawa and the Coulomb potential
for three different laser wavelengths $\lambda$, specified in the panels. 
The color coding represents the instantaneous laser field strength for the time when the electron is at a particular $r$.}
\label{fig:energy}
\end{figure}%
%%%%%%%%%%%%

We conclude with a remark on the so-called ``Coulomb-laser coupling time'' \cite{pana+15,ivsm11,pana+13,scsc+19}\,---\,suggesting that in the case of a Coulomb potential there is a coupling between laser and potential and, conversely, that this is not the case for short-range potentials. 
In general, we have shown that streaking time delays originate from the momentum change $\del{p}$ through the streaking laser. A momentum change of the electron is only possible in the presence of a potential as basic laws of 
energy and momentum conservation forbid energy absorption from light by a free electron \cite{ro95}.
We underline this point with Fig.\,\ref{fig:energy}, which shows the electron's energy as a function of distance $r$
\begin{equation}
E_{\sd=0}(r)=\int\limits_{0}^{t(r)}\!\!\d \tilde{t}\,A(\tilde t)\,V'\big(\x(\tilde{t})-[\xq(\tilde{t})\,{-}\,\xq(0)]\big),
\end{equation}
with the quiver motion $\xq(t)$ in the Kramers-Henneberger picture.
The index ``$\sd{=}0$'' indicates that the electron is released at field maximum, vanishing vector potential or maximum excursion, respectively.
As one can see energy absorption occurs early on, i.\,e., during the first half-cycle of the laser $t\,{<}\,T/2$, cf.\ the respective left-most dark stripes in Fig.\,\ref{fig:energy}, for both short- and long-range potentials.
At the distance traveled in this time by the electron, the asymptotic values of energy absorption (or the asymptotic values of the integrals $\chis$ and $\chic$, respectively) are reached (apart from tiny, quickly-vanishing oscillations) as can be seen in all panels of Fig.~\ref{fig:energy}.
Since for longer wavelengths $\lambda$ also $T/2$ grows, the electron travels further outwards (larger $r$) during this time span, giving rise to more energy absorption, unless this region is already beyond the range of the potential (which is the case for the Yukawa potential in Fig.~\ref{fig:energy}c). Since the Coulomb potential has an infinite range, energy absorption continues to grow with increasing laser period (smaller frequency) as seen in Figs.\,\ref{fig:energy}d--f.

In summary, we have derived the streaking time delays by means of the physically underlying process of energy absorption from the streaking laser which changes the measurable momentum of the photo-electron. 
Our analysis has shown that streaking time delays become independent of the properties of the streaking laser and approach the Wigner-Smith time delay for sufficiently weak streaking fields \emph{and\/} in the limit of vanishing laser frequency or long wavelengths. 

That streaking experiments for atoms measure finite times $\taust$ (despite infinite Wigner-Smith time delays $\del{t}$ for Coulombic systems) is a consequence of finite laser frequencies $\omega$ used. Those finite streaking delays for Coulombic systems, however,
depend on the laser parameters.
In particular they depend on $\omega$ as shown in Fig.\,\ref{fig:coul}a, where $\taust$ 
converges for a Yukawa potential but grows for a Coulomb potential beyond all limits with increasing wavelength (decreasing frequency). 
Unfortunately, the latter even remains true in the case where photo-emission delays from two orbitals in one and the same atom but with different excess energies $E$ are measured simultaneously \cite{scfi+10}.
Since the ``cut-off'' due to a finite $\omega$ depends on the excess energy $E$, it is different for both ionization channels such that the two measured times $\taust$ \emph{cannot\/} be compared directly, see Fig.\,\ref{fig:coul}b.
There, the streaking delays are shown for the Coulomb potential and two energies (dotted and dashed lines in Fig.\,\ref{fig:coul}b).
The resulting relative delay $\Delta\taust$ (black solid) is not well-defined: It depends on the laser frequency/wavelength and is of the same order as the individual times.

Obviously, the interpretation of measured streaking time delays requires careful analysis. 
In particular the last example demonstrates that against any reasonable expectation a relative measurement of a time delay between two orbitals of different binding energy in the same atom and with the same streaking laser does not produce results independent of the properties of the streaking light, namely its frequency. 
However, with the results presented here, it should be possible in the future to design experiments which measure time delays free from properties of the streaking laser.
 
\def\articletitle#1{\textit{#1}.}


\begin{thebibliography}{10}

\bibitem{cota+97}
E. Constant, V.~D. Taranukhin, A. Stolow, and P.~B. Corkum,
  \articletitle{Methods for the measurement of the duration of high-harmonic
  pulses}
Phys. Rev. A {\bf 56},  3870  (1997).

\bibitem{camu+07}
A.~L. Cavalieri, N. M{\"u}ller, T. Uphues, V.~S. Yakovlev, A. Baltu\v{s}ka, B.
  Horvath, B. Schmidt, L. Bl{\"u}mel, R. Holzwarth, S. Hendel, M. Drescher, U.
  Kleineberg, P.~M. Echenique, R. Kienberger, F. Krausz, and U. Heinzmann,
  \articletitle{Attosecond spectroscopy in condensed matter}
Nature {\bf 449},  1029  (2007).

\bibitem{scfi+10}
M. Schultze, M. Fie{\ss}, N. Karpowicz, J. Gagnon, M. Korbman, M. Hofstetter,
  S. Neppl, A.~L. Cavalieri, Y. Komninos, T. Mercouris, C.~A. Nicolaides, R.
  Pazourek, S. Nagele, J. Feist, J. Burgd{\"o}rfer, A.~M. Azzeer, R.
  Ernstorfer, R. Kienberger, U. Kleineberg, E. Goulielmakis, F. Krausz, and
  V.~S. Yakovlev, \articletitle{Delay in photo-emission}
Science {\bf 328},  1658  (2010).

\bibitem{wi55}
E.~P. Wigner, \articletitle{Lower limit for the energy derivative of the
  scattering phase shift}
Phys. Rev. {\bf 98},  145  (1955).

\bibitem{sm60}
F.~T. Smith, \articletitle{Lifetime matrix in collision theory}
Phys. Rev. {\bf 118},  349  (1960).

\bibitem{pana+15}
R. Pazourek, S. Nagele, and J. Burgd{\"o}rfer, \articletitle{Attosecond
  chronoscopy of photo-emission}
Rev. Mod. Phys. {\bf 87},  765  (2015).

\bibitem{dalh+12}
J.~M. Dahlstr{\"o}m, A. {L'Huillier}, and A. Maquet, \articletitle{Introduction
  to attosecond delays in photo-ionization}
J. Phys. B {\bf 45},  183001  (2012).

\bibitem{maca+14}
A. Maquet, J. Caillat, and R. Ta{\"\i}eb, \articletitle{Attosecond delays in
  photo-ionization: time and quantum mechanics}
J. Phys. B {\bf 47},  204004  (2014).

\bibitem{klda+11}
K. Kl{\"u}nder, J.~M. Dahlstr{\"o}m, M. Gisselbrecht, T. Fordell, M. Swoboda,
  D. Gu\'enot, P. Johnsson, J. Caillat, J. Mauritsson, A. Maquet, R.
  Ta{\"\i}eb, and A. {L'Huillier}, \articletitle{Probing single-photon
  ionization on the attosecond time scale}
Phys. Rev. Lett. {\bf 106},  143002  (2011).

\bibitem{dagu+12}
J.~M. Dahlstr{\"o}m, D. Gu{\'e}not, K. Kl{\"u}nder, M. Gisselbrecht, J.
  Mauritsson, A. {L'Huillier}, A. Maquet, and R. Ta{\"\i}eb,
  \articletitle{Theory of attosecond delays in laser-assisted photo-ionization}
Chem. Phys. {\bf 414},  53  (2013).

\bibitem{issq+17}
M. Isinger, R.~J. Squibb, D. Busto, S. Zhong, A. Harth, D. Kroon, S. Nandi,
  C.~L. Arnold, M. Miranda, J.~M. Dahlstr{\"o}m, E. Lindroth, R. Feifel, M.
  Gisselbrecht, and A. L'Huillier, \articletitle{Photo-ionization in the time
  and frequency domain}
Science {\bf 358},  893  (2017).

\bibitem{pana+13}
R. Pazourek, S. Nagele, and J. Burgd{\"o}rfer, \articletitle{Time-resolved
  photo-emission on the attosecond scale: Opportunities and challenges}
Faraday Discuss. {\bf 163},  353  (2013).

\bibitem{ro98}
J.~M. Rost, \articletitle{Semiclassical {S}-matrix theory for atomic
  fragmentation}
Phys. Rep. {\bf 297},  271  (1998).

\bibitem{kigo+04}
R. Kienberger, E. Goulielmakis, M. Uiberacker, A. Baltu\v{s}ka, V. Yakovlev, F.
  Bammer, A. Scrinzi, T. Westerwalbesloh, U. Kleineberg, U. Heinzmann, M.
  Drescher, and F. Krausz, \articletitle{Atomic transient recorder}
Nature {\bf 81},  817  (2004).

\bibitem{itqu+02}
J. Itatani, F. Qu{\'e}r{\'e}, G.~L. Yudin, M.~Y. Ivanov, F. Krausz, and P.~B.
  Corkum, \articletitle{Attosecond streak camera}
Phys. Rev. Lett. {\bf 88},  173903  (2002).

\bibitem{yaba+05}
V.~S. Yakovlev, F. Bammer, and A. Scrinzi, \articletitle{Attosecond streaking
  measurements}
J. Mod. Optics {\bf 52},  395  (2005).

\bibitem{frwi+09}
U. Fr{\"u}hling, M. Wieland, M. Gensch, T. Gebert, B. Sch{\"u}tte, M.
  Krikunova, R. Kalms, F. Budzyn, O. Grimm, J. Rossbach, E. Pl{\"o}njes, and M.
  Drescher, \articletitle{Single-shot terahertz-field-driven {X}-ray streak
  camera}
Nat. Photon. {\bf 3},  523  (2009).

\bibitem{khiv10}
A.~S. Kheifets and I.~A. Ivanov, \articletitle{Delay in atomic
  photo-ionization}
Phys. Rev. Lett. {\bf 105},  233002  (2010).

\bibitem{zhth11}
C.-H. Zhang and U. Thumm, \articletitle{Streaking and {W}igner time delays in
  photo-emission from atoms and surfaces}
Phys. Rev. A {\bf 84},  033401  (2011).

\bibitem{ivsm11}
M. Ivanov and O. Smirnova, \textit{How accurate is the attosecond streak
  camera?}
Phys. Rev. Lett. {\bf 107},  213605  (2011).

\bibitem{suppl}
See supplemental material, deriving the electron’s momentum change
(corresponding to energy absorption) in detail, at [url].

\bibitem{scsc+19}
G. Schmid, K. Schnorr, S. Augustin, S. Meister, H. Lindenblatt, F. Trost, Y.
  Liu, N. Stojanovic, A. Al-Shemmary, T. Golz, R. Treusch, M. Gensch, M.
  K{\"u}bel, L. Foucar, A. Rudenko, J. Ullrich, C.~D. Schr{\"o}ter, T. Pfeifer,
  and R. Moshammer, \articletitle{Terahertz-field-induced time shifts in atomic
  photo-emission}
Phys. Rev. Lett. {\bf 122},  073001  (2019).

\bibitem{ro95}
J.~M. Rost, \articletitle{Analytical total photo cross section for atoms}
J. Phys. B {\bf 28},  L\,601  (1995).

\end{thebibliography}
\end{document}